\documentclass[%
 reprint,
 amsmath,
 amssymb,
 aps,
 nofootinbib,
 prd
]{revtex4-2}

\usepackage{graphicx}% Include figure files
\usepackage{dcolumn}% Align table columns on decimal point
\usepackage{bm}% bold math

\usepackage[T1]{fontenc}
\usepackage[utf8]{inputenc}
\usepackage{amsmath}
\usepackage{mathtools} %added for coloneqq
\usepackage{cancel} %crossing-out terms
\usepackage{MnSymbol} %double angles for classical limit
\usepackage{graphicx,adjustbox} %Lets us rotate symbols 
\usepackage{mathrsfs} % mathscr (for scri+)
\usepackage{mathdots} % triple dots take less space
\usepackage{simpler-wick} %Wick contractions
\usepackage{comment}
\usepackage{autobreak}
\usepackage{relsize}
\allowdisplaybreaks
\numberwithin{equation}{section}

\usepackage{hyperref}
\hypersetup{
	colorlinks=true,
	linktoc=page,
	citecolor=americanrose,
	linkcolor=cadmiumgreen,
	urlcolor=blue} 
\usepackage{xcolor}

\definecolor{americanrose}{rgb}{1.0, 0.01, 0.24}
\definecolor{cadmiumgreen}{rgb}{0.0, 0.42, 0.24}

\makeatletter
\newlength{\apb@width}
\newcommand{\autoparbox}[2][c]{\settowidth{\apb@width}{#2}\parbox[#1]{\apb@width}{#2}}
\newcommand{\includegraphicsbox}[2][]{\autoparbox{\includegraphics[#1]{#2}}}
\makeatletter\g@addto@macro\bfseries{\boldmath}\makeatother

\newcommand{\dd}{\mathrm{d}}
\newcommand{\hd}{\hat{\mathrm{d}}}

\newcommand{\hdelta}{\hat{\delta}}

\newcommand{\sint}{\!\int\!}

\newcommand{\cD}{\mathcal{D}}
\newcommand{\cO}{\mathcal{O}}

\newcommand{\expval}[1]{\langle #1 \rangle}

\newcommand{\oldornew}[1]{}

\begin{document}

\title{\boldmath Weak-field waveforms for generic relativistic orbits}

\author{Stefano De Angelis}
\email{stefano.de-angelis@ipht.fr}

\affiliation{Institut de Physique Théorique, CNRS, CEA, Université Paris-Saclay, F-91191 Gif-sur-Yvette cedex, France}

\date{\today}

\begin{abstract}
    We recast Einstein's equations as ordinary integro-differential equations for the worldlines, integrating out the gravitational field by means of the Schwinger-Keldysh path integral.  
    The same framework allows the gravitational waveform to be computed for unspecified orbits. 
    The two computations are independent: solutions of the equations of motion can then be inserted to reconstruct the waveform for generic orbits.
    The derivation of the equations of motion does not require a map between scattering and bound-orbit observables.
    Thus, it could be implemented within an Effective One-Body-inspired framework, with the advantage that retardation and radiation effects are automatically included: no separation between potential and radiation modes is required. 
    Conversely, the waveform computation may provide an alternative to the Effective One-Body approach, if supplemented by suitable resummation schemes.
    We emphasise that computations in this framework bypass the need for integration-by-parts identities, which are the main technical bottleneck in the computation of observables.
    In this paper, we outline the general framework and present a computational strategy at leading and next-to-leading order in the weak-field expansion.
\end{abstract}

\maketitle

\section{Introduction}
Einstein's field equations are notoriously hard to solve: they form a set of ten coupled, nonlinear partial differential equations, and exact solutions are known only for highly symmetric configurations. 
No closed-form solution is known that describes both the emission of gravitational radiation by generic sources and its back-reaction.
Consequently, waveform modelling for compact-binary coalescences observed by gravitational-wave (GW) detectors relies on approximation schemes or numerical simulations. 
When the separation between the objects is much larger than their Schwarzschild radii, the gravitational interactions are weak. 
This is the typical regime during the inspiral phase of coalescing binaries. 
One can then expand the metric around Minkowski space and solve the equations perturbatively and iteratively. 
This strategy has been successfully applied to a variety of sources and is especially efficient when the source multipoles admit a hierarchical expansion controlled by a small parameter \cite{Thorne:1980ru,Blanchet:1985sp,Blanchet:1987wq,Blanchet:1989ki,Blanchet:1992br}. 
For slowly moving sources, for instance, the expansion parameter is the relative velocity of the constituents. 
In the two-body problem, this is often referred to as the post-Newtonian (PN) expansion; see Ref.~\cite{Blanchet:2013haa} for a thorough review and references therein. 
In the multipolar post-Minkowskian (MPM) formalism, one solves Einstein's equations in the far region outside the source and asymptotically matches the solution to the near region. 
This yields a multipolar expansion of the gravitational wave, in which the radiation multipoles are expressed in terms of the source multipoles. 
The approach separates the problem into two steps: characterising the source and the dynamics of its multipoles, and computing the gravitational waveform once the source multipoles are known. 
In the two-body problem, the source multipoles are expressed in terms of the worldlines of the two bodies for generic trajectories, and the MPM provides a systematic way to compute the waveform and other observables for both elliptic and hyperbolic orbits (\textit{e.g.}, see Refs.~\cite{Bini:2023fiz,Bini:2024rsy,Georgoudis:2024pdz,Bini:2026dvn}). 
The MPM formalism has produced state-of-the-art results for the waveform of quasi-circular binaries \cite{Blanchet:2023bwj,Blanchet:2023sbv}. 
Another well-understood approach is the multipolar Effective Field Theory (EFT) of Refs.~\cite{Goldberger:2009qd,Ross:2012fc}, which describes the interactions of long-wavelength radiation modes with a generic multipolar source. 
When combined with the Non-Relativistic General Relativity (NRGR) framework \cite{Goldberger:2004jt} to describe the microscopic dynamics of the source, this approach is expected to be equivalent to the MPM formalism, up to gauge choices.
Although the EFT framework may be more convenient and systematic, owing to its organisation and the possibility of matching the two regions in a gauge-invariant way, the multipolar EFT has not yet reached frontier-level results for waveform computations. Recent progress has nevertheless been made in this direction (\textit{e.g.}, see Refs.~\cite{Porto:2024cwd,Ivanov:2025ozg}). 
By contrast, state-of-the-art results are available for the dynamics of the system~\cite{Foffa:2016rgu,Foffa:2019rdf,Foffa:2019yfl,Foffa:2019hrb,Blumlein:2021txe,Porto:2024cwd,Brunello:2025gpf,Porto:2026fsd,Brunello:2026anu}, as well as for absorption, tidal, and spin effects.

Much less work has been done for systems that do not admit a hierarchy among multipoles, as is typical in relativistic regimes. 
The main obstacle is technical: for non-relativistic systems, the separation of scales between the dynamics of a system of size $b$ and the radiation wavelength $\lambda\sim\frac{b}{v}$ makes it convenient to split the computation into two parts, which are later matched.
When $\frac{v}{c} \sim 1$, there is no hierarchy between potential and radiation modes. While one may still view this regime as a resummation of an expansion in $\frac{v}{c}$ \cite{Cheung:2018wkq,Bern:2019nnu}, it is sometimes convenient not to expand at all; here we adopt the latter perspective. In this work, we combine the Schwinger-Keldysh path integral \cite{Schwinger:1960qe,Bakshi:1962dv,Bakshi:1963bn,Keldysh:1964ud}\footnote{Given the nature of the problem, where we mainly focus on in-in (inclusive) observables that depend only on the boundary conditions in the past and on the measurement performed in the future, here the waveform, the Schwinger-Keldysh path integral is the natural field-theory formalism. In this context, this point was made for PN EFT in Refs.~\cite{Galley:2009px,Galley:2012hx} and for relativistic scattering in Refs.~\cite{Jakobsen:2022psy,Kalin:2022hph}.} and the worldline post-Minkowskian EFT (PMEFT) \cite{Kalin:2020mvi} to provide a systematic framework. 
Rather than computing observables, such as the scattering angle or the radiated energy, we focus on deriving the equations of motion (EoM), whose solutions are then used to reconstruct the waveform for a generic trajectory. 
In general, we expect analytic solutions of the EoM to be possible only around straight or circular trajectories, although even this is technically non-trivial (\textit{e.g.}, see \cite{Bini:2024hme,Mogull:2025cfn} and references therein). 
A striking lesson from comparing the calculation in Ref.~\cite{Mogull:2025cfn} with the next-to-leading order scattering waveforms computed in Refs.~\cite{Brandhuber:2023hhy,Herderschee:2023fxh,Georgoudis:2023lgf,Elkhidir:2023dco,Bohnenblust:2025gir,Brunello:2025eso} is that most of the complications arise from the trajectory itself, rather than from the structure of the waveform. 
This suggests that, at the same order, the asymptotic radiation field for an unspecified orbit should be relatively simple. 
Such a result should be well suited for numerical evaluation beyond quasi-circular or quasi-straight trajectories. 
This paper presents the general framework and some preliminary results. State-of-the-art results for the EoM and the waveform will be presented in a forthcoming publication.

In the current era of GW science, precision physics requires efficient algorithms for computing accurate gravitational waveforms across the parameter space of binary coalescences. 
This need has pushed researchers to develop new methods for extracting information about the dynamics of the system and the emitted radiation. 
Relativistic perturbation theory has long been used to solve the classical scattering problem in gravity, but on-shell techniques have been applied to gravitational calculations only more recently~\cite{Damour:2016gwp,Damour:2017zjx}. 
Developments in scattering amplitudes~\cite{Bern:1994zx,Bern:1994cg,Bern:1997sc}, together with methods based on a worldline EFT~\cite{Kalin:2020mvi,Mogull:2020sak}, have made possible the calculation of the $G^3$ and $G^4$ corrections, including radiation effects \cite{Bern:2019nnu,Parra-Martinez:2020dzs,Cheung:2020gyp,DiVecchia:2021bdo,Bjerrum-Bohr:2021din,Brandhuber:2021eyq,Kalin:2020fhe,Dlapa:2021npj,Bern:2021dqo,Dlapa:2022lmu,Jakobsen:2023ndj,Jakobsen:2023hig,Damgaard:2023ttc}, while the $G^5$ corrections are currently under way~\cite{Driesse:2024xad,Driesse:2024feo,Bern:2025zno,Bern:2025wyd,Driesse:2026qiz}. 
However, these methods are not suited to the computational strategy presented in this paper because, in the language of the path integral, the worldline variables of the two bodies are integrated out. 
The PMEFT of Ref.~\cite{Kalin:2020mvi} is the only exception.

The push from the scattering-amplitudes community for analytic computations relevant to precision gravitational-wave physics has recently joined a decades-long effort from the General Relativity community, represented by the already-mentioned MPM formalism and the self-force approach (\textit{e.g.}, see Refs.~\cite{Barack:2018yvs,Pound:2021qin} for reviews, and Refs.~\cite{Warburton:2021kwk,Wardell:2021fyy,Mathews:2025txc,Honet:2025lmk,Mathews:2025nyb,Warburton:2025ymy,Barack:2026izc,Geralico:2026kbm} for cutting-edge results and more recent references). These analytic efforts are complemented by various resummation schemes in current incarnations of the Effective One-Body (EOB) approach~\cite{Buonanno:1998gg,Buonanno:2000ef,Pompili:2023tna,Nagar:2023zxh,Albanesi:2025txj,Gamboa:2026jht}.

\section{Framework}

We start by reviewing the essential ingredients of the Schwinger-Keldysh (SK) path integral.
This framework is well suited to computing ``observables'' in classical field theories. 
In classical physics, one usually has a set of EoM that, when supplemented with suitable initial conditions, determine the dynamics of the system both in the future and in the past, provided that sufficient boundary data are specified.\footnote{Interestingly, when the boundary conditions are not fully specified (\textit{e.g.}, in four-dimensional gravitational scattering, where the state is defined at asymptotic infinity), the observables suffer from ambiguities related to BMS transformations. For a recent discussion in the context of waveforms from gravitational scattering, see Refs.~\cite{Veneziano:2022zwh,Georgoudis:2023eke,Bini:2024rsy,Elkhidir:2024izo}.} 
Classical quantities can then be computed as suitable classical limits of expectation values of operators in a state describing the boundary conditions on a hypersurface.\footnote{An alternative to the SK approach, based on squared on-shell S-matrix elements, is the \textit{observable-based} formalism of Ref.~\cite{Kosower:2018adc}, which has also been generalised to radiative observables~\cite{Cristofoli:2021vyo}.}
In the SK formalism, expectation values take the form:
\begin{widetext}
    \begin{equation}
        \expval{\mathcal{O}(x)}_{J^1,J^2} = \sint \cD \phi^1 \cD \phi^2\, \mathcal{O}(x) \, e^{i S[\phi^1] + i \sint \dd^d x\, \phi^1 J^1 - i S[\phi^2] - i \sint \dd^d x\, \phi^2 J^2}\ ,
    \end{equation}
\end{widetext}
where $J^1$ and $J^2$ are sources coupled to the fields $\phi^1$ and $\phi^2$, and the path integral is supplemented by the boundary condition $\phi^1(x) = \phi^2(x)$ at $x^0 \to +\infty$ (or, more generally, at a point in the future of the in state).\footnote{For the systems considered here, no dissipative terms need to be introduced; in the SK formalism, such terms are interactions between the two sets of fields. The energy lost by the binary system is emitted as gravitational waves at infinity.} At this stage, the field $\phi$ is generic; below, we specialise it to the gravitational field and worldline variables. A standard argument shows that changing the field basis is especially convenient for classical computations. We therefore rotate the fields to the Keldysh basis~\cite{Keldysh:1964ud}:
\begin{equation}
    \phi^+ = \frac{\phi^1 + \phi^2}{2}\ ,\quad \phi^- = \phi^1 - \phi^2\ .
\end{equation}
\begin{widetext}
    The classical evolution is obtained retaining the leading order in the advanced  fields~\cite{Caron-Huot:2010fvq}:
        \begin{equation}
            \label{eq:SK_classical_expansion}
            S[\phi^1] + \sint \dd^d x\, \phi^1 J^1 - S[\phi^2] - \sint \dd^d x\, \phi^2 J^2 = \sint \dd^d x\, \left(\frac{\delta S[\phi^+]}{\delta\phi^+} + J^+ \right) \phi^- + \cO (\phi^{-3})\ .
        \end{equation}
\end{widetext}
This simple observation has important consequences \cite{Calzetta:1986ey}:
\begin{enumerate}
    \item In the classical limit, the SK formalism provides a systematic way to compute observables by iterating the equations of motion, restricting the diagrammatic expansion to vertices that involve only one advanced field.
    \item The EoM for the causal field $\phi^+$ are obtained by computing the one-point function of the field $\phi^-$.
\end{enumerate}
In the two-body problem, the dynamical variables of interest are the perturbation of the metric around Minkowski space,
\begin{equation}
    g_{\mu\nu} = \eta_{\mu\nu} + \kappa h_{\mu\nu}\ ,
\end{equation}
and the worldline variables of the two bodies, denoted by $x_i^\mu(\lambda)$, where $\lambda$ is a parameter on the worldline. The action for our problem is
\begin{equation}
    S[x_i,g] = - \sint \dd^d x \left(\frac{2}{\kappa^2} \sqrt{-g} R + \eta^{\mu \nu} F_\mu F_\nu \right) - \sum_{i=1}^{2} m_i \sint \dd \tau_i\ ,
\end{equation}
where the second term is a gauge-fixing term, with $F_{\mu} = \partial^\nu h_{\mu \nu} - \frac{1}{2} \partial_\mu h$, which gives the de Donder gauge. The last term is the point-particle action for the two bodies. For simplicity, we use the reparametrisation invariance of the worldline to choose the proper time $\tau_i$ as the parameter for the minimally coupled worldline:\footnote{In general, one may want to keep the reparametrisation invariance of the worldlines in the final result, but this would give rise to slightly more complicated Feynman rules. We leave this possibility for the future, when state-of-the-art results will be presented using this formalism.}
\begin{equation}
    \label{eq:minimal_coupling}
    - m_i \sint \dd \tau_i \to - \frac{m_i}{2} \sint \dd \tau_i\, g_{\mu \nu} \dot{x}_i^\mu \dot{x}_i^\nu\ ,
\end{equation}
where $\dot{x}_i^\mu = \frac{\dd x_i^\mu}{\dd \tau_i}$.
We now consider the path integral over the gravitational field only, as in the original NRGR framework \cite{Goldberger:2004jt}. Integrating out the gravitational field gives an \textit{in-in} effective action for the worldline variables, which accounts for the gravitational interactions between the two bodies, including retardation and radiation effects:
\begin{equation}
    \label{eq:effective_action}
    \sint \cD h^1_{\mu\nu} \cD h^2_{\mu\nu} e^{i S[x^1_i,\eta+\kappa h^1]-i S[x^2_i,\eta+\kappa h^2]} = e^{i \Gamma[x_i^+,x_i^-]}\ .
\end{equation}
In the classical limit, we expand around small values of the advanced fields:
\begin{equation}
    \Gamma[x_i^+,x_i^-] = \sint \dd \tau_i \sum_{i=1}^{2} \left(m_i \ddot{x}_i^{+ \mu} + \textrm{EoM}^{+ \mu}_i\right) x_{i \mu}^- + \cO ((x_i^-)^3)\ ,
\end{equation} 
where EoM$^+_i$ denotes the interaction terms in the EoM for the worldline variables.
\begin{widetext}
    The waveform can be computed independently as\footnote{One could argue that all the information about classical dynamics is encoded in these two quantities, as higher-point functions factorise \cite{Cristofoli:2021jas}, at least for minimally coupled sources.}
    \begin{equation}
        \label{eq:waveform}
        \expval{h_{\mu\nu}(x)}[x_i^+,x_i^-]\, e^{i \Gamma[x_i^+,x_i^-]} =\! \sint \cD h^1_{\mu\nu} \cD h^2_{\mu\nu}\, h_{\mu \nu}(x) e^{i S[x^1_i,\eta+\kappa h^1]-i S[x^2_i,\eta+\kappa h^2]}\ .
    \end{equation}
    The graviton insertion in the path integral may be either $h^1_{\mu\nu}$ or $h^2_{\mu\nu}$.
    The final integration over the worldline variables localises the waveform on the on-shell trajectories:
    \begin{equation}
        \sint \prod_{i=1}^{2} \cD x_i^+ \cD x_i^-\, \expval{h_{\mu\nu}(x)}[x_i^+,x_i^-]\, e^{i \Gamma[x_i^+,x_i^-]} \simeq \sint \prod_{i=1}^{2} \cD x_i^+\, \delta^{(d)}\!\left(m_i \ddot{x}_i^{+ \mu} + \textrm{EoM}^{+ \mu}_i\right) \expval{h_{\mu\nu}(x)}[x_i^+, 0]\ ,
    \end{equation}
    where we considered the saddle point approximation and focused on the leading term in the advanced worldline variable.
\end{widetext}
The key advantage of this approach is that the two computations in Eqs.~\eqref{eq:effective_action}~and~\eqref{eq:waveform} are independent: 
one can compute connected diagrams with either an advanced worldline source or an external gravitational insertion, with $x_i^- \to 0$ (disconnected diagrams compute the exponential $e^{i \Gamma}$) and expanding in number of retarded-source insertions.
Splitting the problem into the dynamics of the system (including radiation effects) and the formulation of the waveform in terms of generic trajectories resembles the philosophy of the MPM formalism and EOB approaches, and the diagrammatic expansion is a clear simplification compared with attempting a iterative solution of Einstein's equations.
Moreover, the approach separates the genuine gravitational interactions between the two bodies from the contributions associated with the explicit trajectories selected by the boundary conditions. 
This distinction becomes subtle when one computes observables by ``quantising'' the worldline variables or, equivalently, the matter fields in the path integral. 
The standard way to address this issue is to use the map between scattering and bound-orbit observables proposed in Refs.~\cite{Kalin:2019rwq,Kalin:2019inp,Saketh:2021sri,Cho:2021arx}, which maps results from highly eccentric to quasi-circular orbits.
However, this mapping becomes extremely subtle when hereditary effects are included. 
Knowledge of the EoM allows us to bypass this problem.\footnote{As will become clear below, the off-shell computation may be much more subtle, and there remains considerable value in performing simpler computations and finding the correct map to bound orbits at generic eccentricities.}

As in the usual computation of on-shell observables, the perturbative expansion is still organised in powers of $\frac{G m_i}{|x_1 - x_2|}$, and it is useful to consider the following hierarchy of scales:
\begin{equation}
    \label{eq:hierarchy}
    \frac{1}{m_i} \ll G m_i \ll |x_1 - x_2|\ .
\end{equation}
The first inequality is the condition for the validity of the point-particle approximation (\textit{i.e.}, the classical limit), while the second is the condition for the validity of the weak-field expansion. 
The first inequality allows us to discard diagrams with loops of gravitons \cite{Goldberger:2004jt,Holstein:2004dn}. 
Unlike in the scattering setup, we do not assume anything about the angular momentum of the system. 
The fact that the eikonal and trans-Planckian regimes overlap with the classical limit in the large-angular-momentum limit does not mean that the limits are identical. 
In particular, Newtonian mechanics already allows large but bounded scattering angles, as well as closed orbits, while remaining in the weak-field regime. 
An interesting case beyond the weak-field expansion, but still fully classical, is the scattering of a probe particle around a Schwarzschild black hole: in this case, one can also have zoom-whirl orbits \cite{Glampedakis:2002ya}, which are characterised by scattering angles much larger than $2\pi$ \cite{Kol:2021jjc}, with plunges as a limiting case.

\section{The computational setup}

\begin{figure*}[t]
    \centering
    \includegraphics[width=0.125\textwidth]{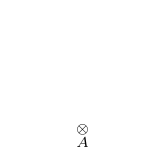}\hfill
    \includegraphics[width=0.125\textwidth]{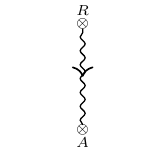}\hfill
    \includegraphics[width=0.125\textwidth]{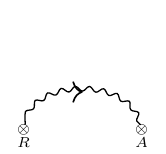}\hfill
    \includegraphics[width=0.125\textwidth]{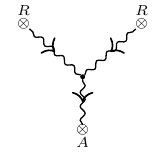}\hfill
    \includegraphics[width=0.125\textwidth]{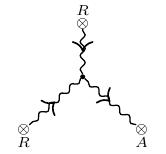}\hfill
    \includegraphics[width=0.125\textwidth]{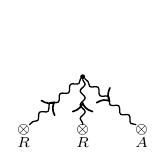}\hfill
    \includegraphics[width=0.125\textwidth]{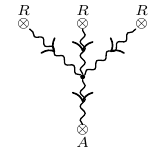}\hfill
    \includegraphics[width=0.125\textwidth]{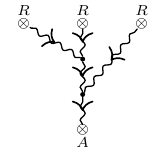}

    \medskip
    \includegraphics[width=0.125\textwidth]{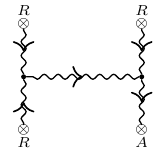}\hfill
    \includegraphics[width=0.125\textwidth]{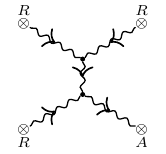}\hfill
    \includegraphics[width=0.125\textwidth]{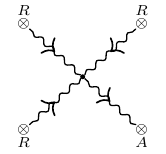}\hfill
    \includegraphics[width=0.125\textwidth]{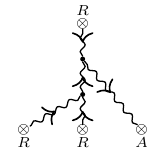}\hfill
    \includegraphics[width=0.125\textwidth]{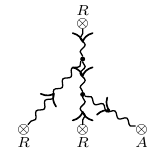}\hfill
    \includegraphics[width=0.125\textwidth]{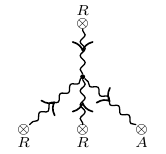}\hfill
    \includegraphics[width=0.125\textwidth]{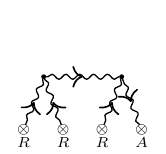}\hfill
    \includegraphics[width=0.125\textwidth]{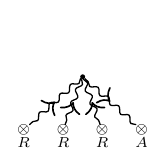}
    \caption{Connected tree-level diagrams contributing to the equations of motion at order $G^L$ with $L=0,\dots ,3$. At the classical level, the expansion in $G$ is an expansion in the retarded source of the gravitational field. The causality flow is indicated by the arrows and is completely determined by the requirement that only one source be advanced. Indeed, the classical limit fixes the momentum flow: there is one outgoing momentum per vertex, as shown in Eq.~\eqref{eq:SK_classical_expansion}. Thus, all propagators with momentum flowing in the direction of the arrows are retarded. The waveform is obtained by summing similar diagrams, but with the advanced insertion removed. Symmetry factors are associated with these diagrams because of the expansion of the action in powers of the retarded sources. In order, the symmetry factors are: $\left\{1;1,1;1/2,1,1/2;1/6,1/2,1,1/2,1/2,1/2,1,1/2,1/2,1/6\right\}$.}
    \label{fig:diagrams}
\end{figure*}

As usual in perturbative computations, we compute the EoM and the waveform as sums of connected tree-level Feynman diagrams. In Fig.~\ref{fig:diagrams}, we list all Feynman diagrams up to $G^3$. The diagrams for the waveform are the same as those for the EoM, up to the insertion of the advanced field. We use dimensional regularisation to regulate divergences, even though no divergences are expected to arise at this order in the perturbative expansion.

\begin{widetext}
    Using the action in Eq.~\eqref{eq:minimal_coupling}, the Feynman rules for the coupling of the worldline variables to the gravitational field take a simple form:\footnote{The symmetrisation over the indices does not require numerical factors.}
    \begin{align}
        \includegraphicsbox{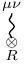} &= - \frac{i \kappa m_i}{2} \sint \dd \tau_i\, \!\int_{\hat{k}} e^{i k \cdot x^+_i(\tau_i)} \dot{x}_i^{+\mu} \dot{x}_i^{+ \nu}\ ,\\
        \includegraphicsbox{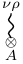} &= \frac{i \kappa m_i}{2} \sint \dd \tau_i\, \!\int_{\hat{k}} e^{i k \cdot x^+_i(\tau_i)} \left(\eta^{\mu (\nu} \ddot{x}_i^{+ \rho)} + i k \cdot \dot{x}_i^{+} \eta^{\mu (\nu} \dot{x}_i^{+ \rho)} - i k^\mu \dot{x}_i^{+ \nu} \dot{x}_i^{+ \rho} \right) x_{i \mu}^{-}\ ,
    \end{align}
\end{widetext}

There are no higher-point worldline couplings, and the only interaction vertices are those of the Einstein-Hilbert action. 
If we were to expand the worldline variable around a straight trajectory and transform to frequency space, we would find the same Feynman rules as in Ref.~\cite{Mogull:2020sak,Kalin:2020mvi}.

Before proceeding, we point out a striking feature of this computation, compared with ordinary perturbative computations of observables: 
integration-by-parts identities~\cite{Tkachov:1981wb,Chetyrkin:1981qh,Laporta:2000dsw,Smirnov:2008iw,Lee:2012cn,Maierhofer:2017gsa,Peraro:2019svx,Wu:2023upw}, which are currently the main technical bottleneck for obtaining precision results, are not needed in this framework. 
Indeed, when we deal with exponential factors, tensor integrals can be computed from scalar integrals simply by acting with derivatives:\footnote{For this observation to work, it was convenient to choose the gauge-fixing term $\frac{1}{\xi} F^\mu F_\mu$ with $\xi = -1$. Any other choice would produce double poles in the graviton momenta, which would have to be reduced or computed separately.}$^,$\footnote{Here, $i$ and $j$ are just labels for the $i^{\rm th}$ integrated graviton momentum and the $j^{\rm th}$ source.}
\begin{equation}
    \label{eq:integrand_reduction}
    k_{i \mu} \to i \frac{\partial}{\partial x^{\mu}_j}\ ,
\end{equation}
where $x_j$, the specific insertion of the worldline, is the Fourier-conjugate variable of $k_i$. Because in our computational setup there are no loops of gravitons, the ``loop'' (integrated) momentum can always be associated with an exponential factor of the external sources.\footnote{This procedure may need refinement at high orders, when UV divergences related to the finite size of the objects start playing a role \cite{Goldberger:2004jt}. We are still very far from this order, and we leave this issue for the future.}

\subsection{The equations of motion}

The free-particle part of the EoM ($G^0$) is obtained from Eq.~\eqref{eq:minimal_coupling} by setting $g_{\mu\nu} \to \eta_{\mu\nu}$ and taking the functional derivative with respect to the advanced worldline variable. This corresponds to the first diagram, with no graviton exchanges, in Fig.~\ref{fig:diagrams}. 

The leading-order interactions are given by two diagrams. Although these are tree-level diagrams, some integrations remain to be performed. After rewriting the rank-one tensor integrals as in Eq.~\eqref{eq:integrand_reduction}, the only integral left to be computed is
\begin{equation}
    G(x_i-x_j) = \int_{\hat{k}} \frac{e^{-i k \cdot (x_i - x_j)}}{k^2 + i \epsilon k^0}\ ,
\end{equation}
which we recognise as the retarded Green's function. In $d$ dimensions, we find
\begin{equation}
    G(x) = - \Theta(x^0) \Theta(x^2)\frac{\left(x^2\right)^{1-\frac{d}{2}}}{2 \pi^{\frac{d-2}{2}} \Gamma \left[2-\frac{d}{2}\right]}\ .
\end{equation}
A surprising, although well-known, observation is that in generic dimensions, and in particular in odd dimensions, the Green's function has support inside the light-cone \cite{Hadamard:1923}. 
In even dimensions, and in particular in four dimensions, the Green's function has support only \textit{on} the light-cone, and intuition, in the form of Huygens' principle, is restored:
\begin{equation}
    \left. G(x) \right|_{d=4} = - \frac{\Theta(x^0)}{2 \pi} \delta (x^2)\ .
\end{equation}
The two diagrams at leading order are qualitatively different: one is the interaction between the two worldlines, while the other is a self-interaction. The former is simple and can be evaluated directly in $d=4$. 
By contrast, the self-interaction is subtle and dimensional regularisation\footnote{Or any other suitable regulator.} is needed to regulate its divergences. 
In $d$ dimensions, the Green's function has support inside the light-cone, and the entire past history of the worldline lies inside this region. 
A similar computation was performed in Ref.~\cite{Galley:2010es} for electrodynamics, where it was shown that this self-interaction gives rise to the Abraham-Lorentz-Dirac force.
\begin{widetext}
    From the first diagram, we obtain a Fokker-Wheeler-Feynman-type interaction between the two worldlines~\cite{Fokker:1929utg,Wheeler:1949hn,Friedman:2005rx}:
    \begin{align}
        \includegraphicsbox{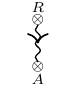} &= i \sint \dd\tau\, x_{1 \mu}^- \left\{4 G m_1 m_2\!\int_{-\infty}^{\tau}\! \dd t \left[\left(\ddot{x}_1^{+\mu} - 2 \ddot{x}_1^{+}\cdot \dot{x}_2^{+} \dot{x}_2^{+ \mu}\right)\delta((x^+_1 - x^+_2)^2) + \right. \right.\\
        &\notag + \left(-4 \dot{x}^+_1\cdot x^+_1 %\left(\dot{x}_1^{+ \mu } -2 
        \dot{x}^+_1\cdot \dot{x}^+_2 \dot{x}_2^{+ \mu }
        %\right)
        - 2 \dot{x}^+_1\cdot x^+_2 \left(\dot{x}_1^{+ \mu }-2 \dot{x}^+_1\cdot \dot{x}^+_2 \dot{x}_2^{+ \mu }\right)+\left(2 (\dot{x}^+_1\cdot \dot{x}^+_2)^2-1\right)\left(x_1^{+ \mu }-x_2^{+ \mu }\right)\right)\delta^\prime((x^+_1 - x^+_2)^2)\big] \Big\}
    \end{align}
    where all worldline variables of particle 1 are evaluated at time $\tau$ and all worldline variables of particle 2 are evaluated at time $t$.
\end{widetext}
The delta functions can be further simplified. The radiation-reaction diagram requires more care: one could repeat the previous computation after substituting $x_2^{+ \mu}(t)\to x_1^{+\mu}(t)$ and find a divergence at $t=\tau$. To regulate this divergence, we perform the computation in dimensional regularisation. Away from $d=4$, the causal structure of the Green's function is different, and the support lies within the (past) light-cone. Since the worldline is time-like, its entire past history lies within this region. On the other hand, any finite contribution from the integration of the convolution of the Green's function with the trajectory gives a contribution of $\cO (d-4)$. Thus, we can simply replace the integration along the entire trajectory with an integral around the coincidence point:
\begin{equation}
    \int_{-\infty}^{\tau}\dd t\, f(t) = \int_{-\epsilon}^{\tau}\dd t\, f(t) +\cO (d-4)\ ,
\end{equation}
and expand the trajectory:
\begin{equation}
    \begin{split}
        \label{eq:trajectory_expansion}
        x_1^{+ \mu}(t) = x_1^{+ \mu}(\tau) & + \dot{x}^{+ \mu}(\tau) (t-\tau) + \frac{1}{2} \ddot{x}^{+ \mu}(\tau) (t-\tau)^2\\
        & + \frac{1}{3!} \dddot{x}^{+ \mu}(\tau) (t-\tau)^3+\dots \ ,
    \end{split}
\end{equation}
where higher-order terms give convergent integrals that are suppressed in the four-dimensional limit.
\begin{widetext}
    Performing the time integration and taking the limit $d\to 4$, we find
    \begin{equation}
        \includegraphicsbox{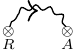} = i \sint \dd\tau\, x_{1 \mu}^- \left\{G m_1^2 \left[\left(2 (\dot{x}_1\cdot \ddot{x}_1)^2-3 \ddot{x}_1^2+\frac{2}{3}\dot{x}_1\cdot \dddot{x}_1\right) \dot{x}_1^{\mu} -\frac{11}{3} \dddot{x}_1^{\mu }\right]\right\}
    \end{equation}
    where, for convenience, we have set $\dot{x}_1^2 = 1 + \cO(G)$. The first two terms in Eq.~\eqref{eq:trajectory_expansion} give scaleless integrals.

    % A post-Newtonian expansion of the integrand and the integral produces the usual instantaneous Newtonian potential, plus an infinite series of corrections.

    At order $G^2$, there are three diagrams, but again only one integral must be computed. 
    This integral can be computed as a convolution of three Green's functions:
    \begin{equation}
        \label{eq:convolution}
        G(x_i,x_j;x) = \int_{\hat{k}_1,\hat{k}_2} \left. \frac{e^{i x_i \cdot k_1 + i x_j \cdot k_2 - i x \cdot k}}{(k_1^2 + i \epsilon k_1^0)(k_2^2 + i \epsilon k_2^0)(k^2 + i \epsilon k^0)}\right|_{k=k_1+k_2} = \int_y G(y-x_i) G(y-x_j) G(x-y)\ .
    \end{equation}
    This representation makes it manifest that the seemingly complicated integrals with exponential factors are simply convolutions of Green's functions, as naturally expected from the iteration of classical EoM. 
    At this order, the computation of this integral (and its derivatives) in $d=4$ is relatively simple (for example, by shifting the integration variable $y \to y + x$):
    \begin{equation}
        G(x_i,x_j;x) = \frac{1}{32 \pi^3 \sqrt{|\Delta |}} \sint \dd^2\!\rho\, \Theta(y^0-y_i^0) \Theta(y^0-y_j^0) \Theta(-y^0)\, \delta\left(\rho^2 + \frac{y_i^2 y_j^2 y_{ij}^2}{4 \Delta}\right)\ ,
    \end{equation}
\end{widetext}
where $\Delta = y_i^2 y_j^2 - (y_i \cdot y_j)^2$, $y_i = x_i - x$, and $y_{ij} = x_i - x_j$. The integration region has been restricted by two of the three $\delta$ functions to
\begin{equation}
    y^\mu = - \frac{(y_i^2 - y_i \cdot y_j) y_j^2}{2 \Delta} y_i^\mu + (i\leftrightarrow j) + \rho^\mu\ ,
\end{equation}
with $\rho \cdot y_i = \rho \cdot y_j = 0$. 
If either $y_i^\mu$ or $y_j^\mu$ is time-like, the integration over $\rho$ is trivial and gives a $\pi$ factor. 
This is the case, for example, if we are computing the EoM for a worldline and one of the sources is the worldline itself. 
If both sources belong to the other worldline, we can repeat the same computation as before, but with the shift $y \to y + x_i$. 
The linear combinations in the $\Theta$ functions will change. 
In the former case, it is easy to see that no singularity arises from the integration over the proper time of the worldline as the sources get closer to the insertion of $x_i^{- \mu}(\tau_i)$.

At higher orders, the integrals are better computed using differential equations~\cite{Kotikov:1990kg,Bern:1993kr,Remiddi:1997ny,Gehrmann:1999as,Henn:2013pwa,Henn:2014qga}. 

The equations of motion at order $G^2$ have already been computed in Refs.~\cite{Bel:1981be,Westpfahl:1979gu}. The precise matching with these results, which may involve subtleties related to gauge and regularisation choices, is beyond the scope of this paper. 

\subsection{The waveform}

The computation of the waveform at order $G^L$ proceeds in parallel with that of the EoM at order $G^{L+1}$, but with an important simplification: the insertion of the advanced field is in the bulk of spacetime, while the waveform is evaluated at asymptotic (null) infinity. In practice, the waveform at order $G^L$ can be obtained by simple manipulations of the EoM diagrams at order $G^{L+1}$ and by taking a limit of the integrals. There are two strategies for performing these integrals: taking the limit of the bulk integrals needed for the EoM, as discussed above, or performing the asymptotic expansion of Eq.~\eqref{eq:convolution} at $t,r\to \infty$ with $u = t-r$ fixed, before integration.\footnote{At higher orders, when infrared divergences arise, the former will correspond to an asymptotic expansion of the integrals with $\frac{\log^m r}{r}$ corrections to the leading $1/r$ behaviour. The latter strategy will require an infrared regulator, such as dimensional regularisation, and we will find poles in $\frac{1}{d-4}$. Both contributions resum to an exponential that amounts to a constant shift of the retarded time \cite{Weinberg:1965nx,Blanchet:1987wq,Porto:2012as,Brandhuber:2023hhy,Herderschee:2023fxh,Georgoudis:2023eke,Caron-Huot:2023ikn}.}
We follow the second strategy here. We consider the future null infinity limit
\begin{equation}
    x^\mu = r n^\mu + \cO(r^0)\ , \quad x^2 = 2 r u\ ,
\end{equation}
where $n^\mu$ is a null vector. In the limit $r \to \infty$, the Green's function simplifies to
\begin{equation}
    G(x-y) \simeq - \frac{1}{4 \pi r} \delta (u - n \cdot y)\ ,
\end{equation}
and similarly for the $d$-dimensional case, which is not needed for the present computations. The leading-order contribution to the waveform is 
\begin{equation}
    \epsilon_{\mu \nu} h^{\mu \nu}(u) = \sum_{i=1}^{2} \frac{4 G m_i}{r} \sint \dd \tau_i\, \delta(u - n \cdot x_i^+(\tau_i)) \epsilon_{\mu \nu} \dot{x}_i^{+\mu} \dot{x}_i^{+\nu}\ ,
\end{equation}
where the delta function is easily resolved using an explicit parametrisation of the worldline. The integral at the next order is a simplified version of the convolution in Eq.~\eqref{eq:convolution}, with one of the Green's functions replaced by its asymptotic form:
\begin{widetext}
    \begin{equation}
        \begin{split}
            G(x_i,x_j;r n) & = - \frac{1}{16 \pi^3 r} \!\int_y \Theta(y^0-x_i^0) \Theta(y^0-x_j^0) \delta\left((y-x_i)^2\right) \delta\left((y-x_j)^2\right) \delta(u - n \cdot y)\ \\
            & = \frac{\Theta\left(u-\frac{1}{2} X_{i j} \cdot n\right) \Theta\left(-x_{i j}^2\right)}{32\pi^2 r \sqrt{(n\cdot x_{i j})^2}}\ ,
        \end{split}
    \end{equation}
\end{widetext}
where $X_{i j}^\mu = x_i^\mu + x_j^\mu$ and $x_{i j}^\mu = x_i^\mu - x_j^\mu$.
Similar computations, specialised to scattering trajectories, was performed long ago in Refs.~\cite{Kovacs:1977uw,Kovacs:1978eu}. Recently, this result was rederived using on-shell worldline\footnote{In the spirit of this paper, the worldline in Refs.~\cite{Mogull:2020sak,Jakobsen:2021smu} is integrated over (``quantised''), in contrast to the PMEFT approach.} and amplitude methods in Refs.~\cite{Jakobsen:2021smu,Mougiakakos:2021ckm,DeAngelis:2023lvf}. 
The simplicity of this integral is remarkable, if one considers the complexity of the integrals in the on-shell approach already at leading order\cite{Mougiakakos:2021ckm,Brunello:2025eso}.

\section{Conclusions and outlook}
In this paper, we introduced a strategy to simplify the binary problem in the weak-field regime for generic relativistic scenarios, by computing the EoM for the worldlines and the waveform as independent quantities using the SK path integral. 

This formalism is well suited to computations at higher orders in the weak-field expansion and does not require the use of integration-by-parts identities, which are currently the main technical bottleneck for obtaining precision results. 
On the other hand, the integrals that appear in this framework are slightly more complicated than the usual Feynman integrals, because they involve exponential factors in the coupling of the gravitational field to the worldlines. 
Nonetheless, they can be recast as convolutions of Green's functions in momentum space, which resemble the iteration of classical EoM. 
From a conceptual point of view, the SK path integral provides a systematic and diagrammatic way to iterate Einstein's equations for the gravitational field.
We illustrate the simplicity of this approach by computing the leading-order terms in the EoM and the waveform, and by displaying the type of integrals that appear at next-to-leading order.
We will present the state-of-the-art results, the comparison with the non-relativistic results (emphasising the new structural features of the method), and the detailed computations of the waveform in a companion paper.

The new strategy amounts to abandoning the idea of obtaining analytic closed-form solutions for observables or trajectories.
Instead, we focus on deriving analytically only the EoM for two (or more) gravitationally interacting worldlines and the asymptotic gravitational field for generic trajectories.
These are the two basic ingredients for numerically simulating the relativistic dynamics of a binary system in the weak-field regime and recording the corresponding waveform. 
Even in Newtonian gravity, the existence of analytic solutions over the entire phase space is a remarkable feature, tied to the integrability properties of Newtonian mechanics. As one changes the number of dimensions or adds the first post-Newtonian correction, integrability generically breaks down, and only approximate solutions remain available, or analytic solutions around the simplest trajectories.\footnote{For example, one should not expect an analytic description of some of the orbits that future observations are expected to probe.} 
On the other hand, if an analytic solution for the trajectories is known in the appropriate gauge, the waveform can also be obtained analytically.

This approach is similar to the EOB approach, but the latter is based on post-Newtonian intuition: the EOB relies on a conservative Hamiltonian description of the dynamics, supplemented with information about radiation reaction and the fluxes at infinity.
Unlike in the EOB approach, no resummation scheme has yet been proposed for the EoM and the waveform. Thus, the only resummed information lies in the trajectories, which are not expanded around straight lines (constant velocity) or circular orbits (constant frequency).
A natural future direction is to combine ideas from the two approaches, for instance by translating some of the resummation schemes of the EOB approach into the framework presented here.

The general framework presented here makes the computation of higher-order corrections to the EoM and the waveform systematic within the SK and worldline formalisms. 
As mentioned before, the integrals at higher orders are expected to be slightly more complicated. 
However, the fact that integration-by-parts identities are not needed is a substantial advantage, and we expect it to make the computation of higher-order corrections much more efficient. 
Nevertheless, integration-by-parts identities have been used in this context to reduce integrals with exponential factors~\cite{Brunello:2023fef,Brunello:2024ibk,Brunello:2025cot,Brunello:2025eso}. 
In general, we expect explicit computations in this framework to be much simpler than those of Ref.~\cite{Brunello:2025eso}, because we work directly in position space, where the iterated Bessel functions~\cite{Bini:2024ijq,Gaiur:2026} do not appear.

Another interesting direction would be to evaluate the convolutions of the Green's functions in the EoM and the waveform integrands using the \textit{method of regions}~\cite{Beneke:1997zp}. 
Indeed, we know from the study of the action that only \textit{potential} and \textit{radiation} modes contribute to the final result. 
The method of regions allows one to systematically expand the integrand in the relevant regions, thereby drastically simplifying the integrals. 
This method could be an alternative to the PN EFT approach, because it bypasses the need for splitting into the potential region and the radiation region (multipolar EFT).
This approach simplifies the computation, but it loses some of the advantages of working with two EFTs.\footnote{In EFTs, it is always useful to split computations into different regions, because running and matching between them allow one to resum large logarithms that would not be resummed in a unified setup.}
For example, this splitting allows one to resum universal contributions in the UV of the multipolar EFT, before matching it to the source-region computation~\cite{Goldberger:2009qd,Goldberger:2012kf}.
In any case, this approach is useful because it would bring the framework into closer contact with EOB methods.
In parallel, one could try to make closer contact with the self-force literature, for example by replacing the PMEFT with the EFT for Extreme Mass Ratios of Refs.~\cite{Cheung:2023lnj,Cheung:2024byb}.
In particular, we would like to emphasise some technical similarities between the framework presented in this paper and the Green's function approach of Ref.~\cite{Wardell:2014kea}.

Finally, an interesting direction is to explore whether the EoM and waveform integrands can be computed using recursive \cite{Berends:1987me} and/or unitarity-based methods \cite{Bern:1994zx,Bern:1994cg}, in the spirit of Refs.~\cite{Edison:2023qvg,Edison:2024owb,Jakobsen:2023oow,Haddad:2025cmw,He:2025how}.

\section*{Notation} 
As is standard in QFT, we work in the mostly minus signature $(+,-,\dots,-)$ and use natural units $c=1$ and $\hbar=1$ (unless otherwise stated). The gravitational coupling is $\kappa = \sqrt{32 \pi G}$, where $G$ is Newton's constant. We adopt the $\pi$-absorbing \textit{hat} notation~\cite{Kosower:2018adc} for the integral measure and delta functions,
\begin{equation}
\hd^d k := \frac{\dd^d k}{(2\pi)^d}\ , \quad  \hdelta^{(d)}(\cdot ):= (2 \pi)^d \delta^{(d)}(\cdot)\ .
\end{equation}
We also introduce the shorthand notation
\begin{equation}
\int_{\hat{k}} = \int \hd^d k\ , \quad \int_{x} = \int \dd^d x\ .
\end{equation}

\acknowledgments
I would like to thank Brando Bellazzini, Giacomo Brunello, David Kosower, Marcello Romano, Nikola Savic, and Filippo Vernizzi for useful discussions. I also thank Donato Bini, Thibault Damour, Jan Steinhoff, Radu Roiban, and Fei Teng for collaboration on related topics. I am especially grateful to Brando Bellazzini, Donato Bini, Giacomo Brunello, Carlo Heissenberg, David Kosower, Jan Steinhoff, Radu Roiban, and Marcello Romano for comments on the draft, and to Davide Usseglio for helping me navigate the self-force literature. The Feynman rules were computed using the \texttt{Mathematica} package \href{https://www.xact.es}{\texttt{xAct}}. AI tools were used extensively for coding, formatting the manuscript, research, and writing, alongside many other tools, software, and online resources that are usually not explicitly mentioned in the acknowledgements. 

\bibliographystyle{apsrev4-1}
\bibliography{binary.bib}

% \onecolumngrid
% \newpage
% \appendix

\end{document}